# Low-temperature mean valence of nickel ions in pressurized $La_3Ni_2O_7$

Shu Cai[1]*, Yazhou Zhou[2]*, Hualei Sun[3]*, Kai Zhang[1]*, Jinyu Zhao[1], Mengwu Huo[3], Lucie Nataf[4], Yuxin Wang[2], Jie Li[5], Jing Guo[2], Kun Jiang[2], Meng Wang[6], Yang Ding[1], Wenge Yang[1], Yi Lu[5,7], Qingyu Kong[4]†, Qi Wu[2], Jiangping Hu[2,8]†, Tao Xiang[2,9], Ho-kwang Mao[1] and Liling Sun[1,2]†

[1] *Center for High Pressure Science & Technology Advanced Research, 100094 Beijing, China*
[2] *Institute of Physics, Chinese Academy of Sciences, Beijing 100190, China*
[3] *School of Science, Sun Yat-Sen University, Shenzhen, Guangdong 518107, China*
[4] *Synchrotron SOLEIL, L'Orme des Merisiers, Saint-Aubin-BP48, 91192 Gif-sur-Yvette Cedex, France*
[5] *National Laboratory of Solid State Microstructures and Department of Physics, Nanjing University, Nanjing 210093, China*
[6] *School of Physics, SunYat-Sen University, Guangzhou, China.*
[7] *Collaborative Innovation Center of Advanced Microstructures, Nanjing University, Nanjing 210093, China*
[8] *New Cornerstone Science Lab, Beijing 100190, China*
[9] *Beijing Academy of Quantum Information Sciences, Beijing 100193, China*

The discovery of high critical temperature ($T_c$) superconductivity in pressurized $La_3Ni_2O_7$ has ignited renewed excitement in the search of novel high-$T_c$ superconducting compounds with $3d$ transition metals. Compared to other ambient-pressure superconductors, such as copper-oxide and iron-oxypnictides, unraveling the mechanisms of the pressure-induced superconductivity poses significant and unique challenges. A critical factor in this phenomenon seems to be related to the electronic configuration of $3d$ orbitals, which may play a fundamental role in driving high-$T_c$ superconductivity. However, the pressure effects on the mixed-valence states of $3d$-orbital cations and their influence on the emergence of high-$T_c$ superconductivity remain poorly understood. Here, we use high-pressure ($P$) and low-temperature synchrotron X-ray absorption spectroscopy to investigate the influence of pressure on the mean valence ($\nu$) change of Ni ions in $La_3Ni_2O_7$. Our results demonstrate that at a low-temperature of 20 K, $\nu$ remains relatively stable across the pressures range from 1 atm to 40 GPa. Based on analyzing the absorption data, we find that, at a critical pressure, the ambient-pressure ordered phases disappear and both the structural and the superconducting phase transition occur. The pressure-induced structural phase transition revealed by our absorption results is consistent with that determined by X-ray diffraction, offering new information for a comprehensive understanding on the pressure-induced superconductivity in $La_3Ni_2O_7$.

Recently, signatures of superconductivity near 80 K were observed in the $La_3Ni_2O_7$ single crystal under pressure [1], manifesting the discovery of a novel high-$T_c$ superconductor among $3d$ transition-metal compounds, after the findings of copper-oxide and Fe-based superconductors at ambient pressure [2-15]. Subsequently, some experimental efforts [16-30] and a lot of theoretical input [19, 31-44] has been dedicated to understand the superconducting properties and mechanisms, with the attempt to identify the common essential ingredients that contribute to the high-$T_c$ superconductivity in $La_3Ni_2O_7$, iron pnictide and the copper oxide superconductors with different $3d$ transition metals. Experimental results have shown that $La_3Ni_2O_7$ possesses a non-superconducting orthorhombic structure at ambient pressure. In particular, Ni ions exhibit a mean valence of $2.5^+$ [1, 11, 15, 18, 20, 22, 23, 33, 41-43, 45], leading to the coexistence of spin-density-wave (SDW) and density wave (DW) ground states [12, 13, 18, 21, 46, 47]. This indicates the vital role played by the mixed valence of Ni ions in shaping the electronic state and transport properties of this material. Under high pressure conditions, the transition from orthorhombic to tetragonal phases occurs at a critical pressure ($P_c$) and then the filamentary-like superconductivity emerges from the new phase [6, 10, 48]. Therefore, one of the key issues is naturally raised: are there any changes in the valence state of Ni ions under pressure? If so, it may suggest that the change in mean valence may play a crucial role for the structural alterations and the corresponding superconducting transition. Alternatively, if not, it would imply that the pressure-induced phase transition governs the change of the electronic state of the system.

Under ambient-pressure and room-temperature conditions, the utilization of X-ray absorption spectroscopy (XAS) has demonstrated its effectiveness in examining the electronic configurations of cations in transition-metal compounds [49-52]. By integrating this method with high-pressure and low-temperature conditions to analyze the electronic and crystal structures of $La_3Ni_2O_7$, with a specific focus on the pressure-induced change in charge configuration of Ni ions, some new insights into the microscopic physics of the high-$T_c$ superconductivity in this compound may be provided. In this study, we performed the XAS investigations at the Ni $K$-edge under high-pressure and low-temperature conditions for both the single crystal and polycrystalline $La_3Ni_2O_7$ to unveil the connection between the mean valence of Ni ions with the structural change and the superconducting transition in the pressurized $La_3Ni_2O_7$.

The single crystal $La_3Ni_2O_7$ was grown using a vertical optical-image floating zone furnace, the detailed information can be found in Ref. [1], and the polycrystalline $La_3Ni_2O_7$ was synthesized using the sol-gel method [17, 48]. A high-pressure environment was generated using a diamond anvil cell made of BeCu alloy. Diamond anvils with 300 μm culets (flat area of the diamond anvil) were used for independent measurements. The sizes of the gasket hole and the sample are 150 μm and 100 μm, respectively. To maintain the sample in a quasi-hydrostatic pressure environment, we employed silicone oil as the pressure transmitting medium. Before closing the diamond anvils, a ruby flake was placed on the top of the sample to ensure that the ruby experiences the same pressure as the sample. Through the window of the cryostat, we are able to determine the pressure of the sample at 20 K by using the ruby fluorescence method [53]. Our XAS measurements were conducted at the ODE beamline at Soleil synchrotron. The incident fluxes were around $10^{12}$ photons/second with a beam size of approximately 100 μm. Energy calibration was performed using a Ni foil, and the data were normalized based on the incident flux. XAS measurements were conducted across an energy span of 200 eV with a resolution of 0.5 eV using the Si311 monochromator.

First, we conducted XRD diffraction measurements on both the single crystal and polycrystalline samples at ambient conditions, confirming the samples crystallize in an orthorhombic unit cell in the *Cmcm* space group (Fig.1a-1d). Our structural refinement results revealed that the lattice parameters of the samples are $a = 20.465$ Å, $b=5.408$ Å and $c =5.453$ Å, in agreement with results reported in Ref. [1, 17].

Subsequently, we conducted X-ray absorption spectroscopy measurements at the Ni *K*-edge for the $La_3Ni_2O_7$ samples under high-pressure and low-temperature conditions. The spectra collected at 20 K for both the single crystal and polycrystalline samples are shown in Fig.2. Significantly, we observed the shifts in the white line peak (labeled A) towards higher energies in both the single crystal and polycrystalline samples as pressure increases (Fig.2a and 2c). Specifically, the white line peak of the single crystal sample shifts to higher energy by 0.61 eV at 21 GPa (Fig.2a). Similarly, the corresponding peak of the polycrystalline sample shifts by 0.57 eV at 19.5 GPa (Fig.2c), reflecting that the observation from the polycrystalline sample is consistent with that from the single crystal sample at the same pressure level. These results are the direct evidence of the contraction of the distance between Ni and its neighboring atoms

[54, 55], which is also closely associated with the changes of the Ni ion configuration if the change occurs.

To provide a clear representation of the variations in X-ray absorption edges under pressure, we present the first derivatives of the XAS data collected at different pressures (Fig. 2b and 2d). Our analyses consistently demonstrated that these peaks shift to higher energies. It is well known that two factors influence the energy shift of the white line peak: one corresponds to lattice contraction, and the other is linked to valence alteration. To identify the origin of the pressure-induced energy shift observed in $La_3Ni_2O_7$, we examine the energy shifts as a function of the contraction of the Ni-O bond for $La_3Ni_2O_7$ ($Ni^{2.5+}$), $YNiO_3$ ($Ni^{3+}$) and $NiO$ ($Ni^{2+}$) through analyzing the experimental results. As illustrated in Fig.3, the rate of change in energy shift versus Ni-O bond is nearly the same for $La_3Ni_2O_7$, $YNiO_3$ and $NiO$, suggesting that the observed energy shift in $La_3Ni_2O_7$ is predominantly influenced by the lattice contraction rather than the valence alteration. To further confirm the results mentioned above, we performed the theoretical calculations for $La_3Ni_2O_7$ ($Ni^{2.5+}$), $LaNiO_3$ ($Ni^{3+}$) and $NiO$ ($Ni^{2+}$) and obtained the same conclusion (the details of the calculations can be found in Ref. [56], Supplementary Information-SI). Our results agree with the experimental results reported very recently [29]. Further investigation is warranted to understand why the application of pressure does not induce a change in the valence state of Ni ions in $La_3Ni_2O_7$.

We summarize the pressure dependence of the mean valence (ν) of Ni ions in Fig.4a, showing that ν remains almost unchanged in the pressure range from 1 atm to 40 GPa. Details of determining the mean valence (ν) of Ni ions can be found in Ref.

[56]. Moreover, since the changes in the integrated area of the white line peaks are indicative of structural information[57, 58], we investigated the effect of pressure on the crystal structure by analyzing the integral area of the white line peak measured from both the single crystal and polycrystalline samples at different pressures. As shown in Fig.4b, the area monotonously decreases as pressure rises, but its slope dramatically alters at the critical pressure of $P_{c1}$ ($P_{c1}$ = 12.5 GPa, which is estimated through the average pressure of (11 GPa+14 GPa)/2)), indicating a structural phase transition. This result aligns with the findings of high-pressure XRD studies on the single crystal sample with the same pressure calibrating method of ruby fluorescence [6]. Based on these results, we propose that the lattice stability and the emergence of superconductivity in pressurized of $La_3Ni_2O_7$ is not related to the mean valence state of Ni ions. To demonstrate the pressure-induced co-evolutions among the mean valence of Ni ions, lattice structure and superconductivity in $La_3Ni_2O_7$, we plotted its pressure-temperature phase diagram in Fig. 4c. Given that the Ni ions exhibit a mixed valence of $2.5^+$ at ambient pressure, it implies the coexistence of Ni ions in both $2^+$ and $3^+$ states in the lattice. While, as pressure increases, the mean valence (ν) of Ni ions remains nearly unchanged. However, at a critical pressure $P_{c1}$ (12.5 GPa), structural and corresponding superconducting phase transitions occur, where the DW and SDW phases disappear. This observation allows us to propose that the pressure-induced structure phase transition is crucial to cease the presence the DW and SDW phases and then boost the superconducting transition in the pressurized $La_3Ni_2O_7$.

In conclusion, we report the experimental results, achieved through high-pressure and low-temperature XAS measurements, on the pressure effect on the mean valence of Ni ions in the single crystal and polycrystalline $La_3Ni_2O_7$. Based on the observations, we find that mean valence of Ni ions $La_3Ni_2O_7$ remains almost constant under pressure within the range from 1atm to 40 GPa. but the pressure dependent area of the while line peak changes its slope at 12.5 GPa, where the DW and SDW states are fully suppressed and the superconducting transition occurs. Our study excludes the possibility that mean valence change of the Ni ions is crucial for the emergence of superconductivity, instead, reveal that the pressure-induced structure change plays a fundamental role in ceasing the DW and SDW states, thus in triggering the superconducting phase transition.

These authors with star (*) contributed equally to this work.
Correspondence and requests for materials should be addressed to Qingyu Kong (kong@synchrotron-soleil.fr), Jiangping Hu (jphu@iphy.ac.cn) and Liling Sun (llsun@iphy.ac.cn or liling.sun@hpstar.ac.cn)

**Acknowledgements**

The work was supported by the NSF of China (Grants No. 12494590, U2032214, 12122414, 12104487, 12004419, 12474054, 12425404, 12474137 and 12274207) and the National Key Research and Development Program of China (Grants No. 2022YFA1403900, 2021YFA1401800, 2023YFA1406000, 2023YFA1406500 and 2022YFA1403000). Partial work was supported by the Synergetic Extreme Condition User Facility (SECUF).

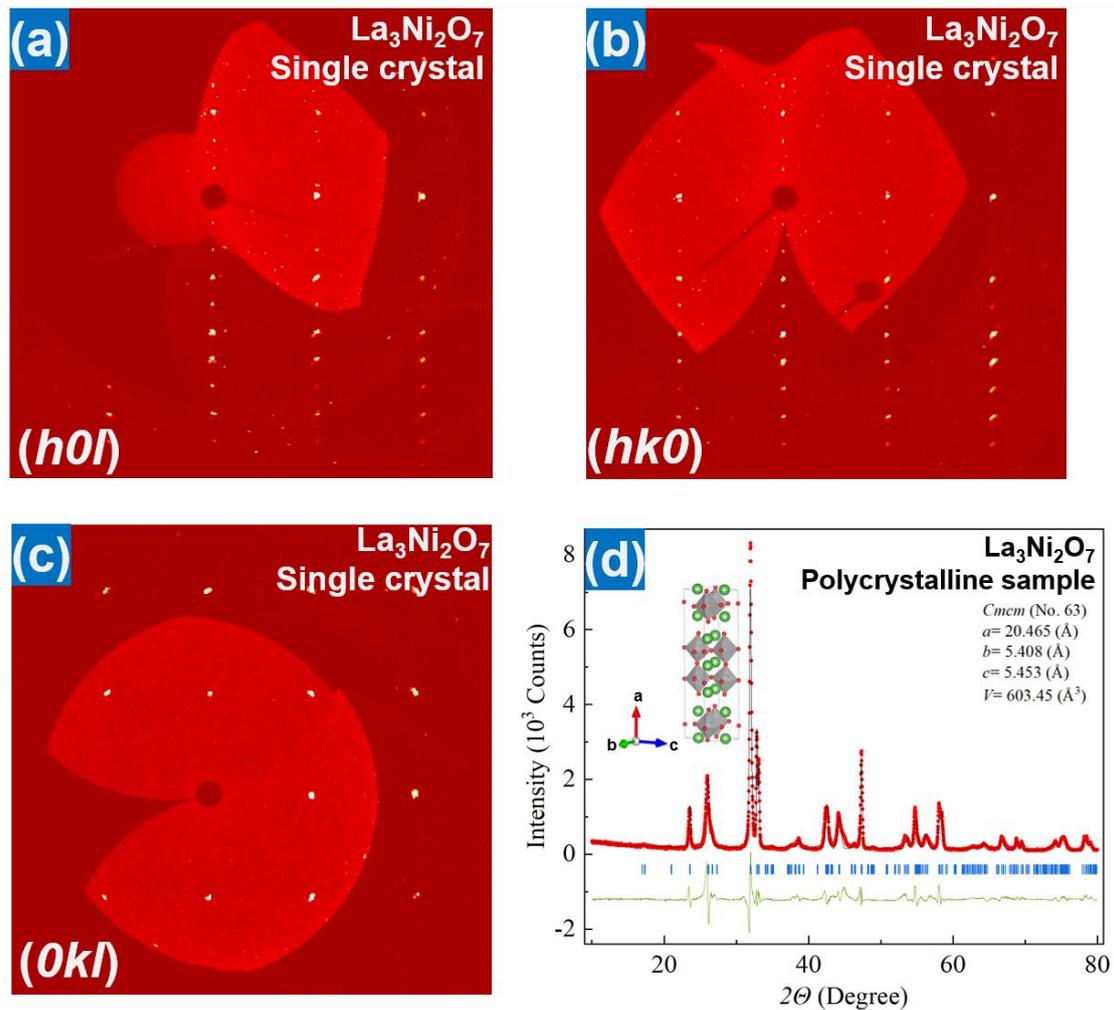

**Figure 1. Structural characterization for the single crystal and polycrystalline La$_3$Ni$_2$O$_7$.** (a)-(c) Single crystaX-ray diffraction patterns for the (*0kl*), (*h0l*) and (*hk0*) zones taken at ambient pressure and 300 K for the single crystal La$_3$Ni$_2$O$_7$. (d) Powder X-ray diffraction pattern collected at ambient condition for the polycrystalline La$_3$Ni$_2$O$_7$ (the X-ray wavelength is 1.54056 Å). All data show that the sample possesses the *Cmcm* phase.

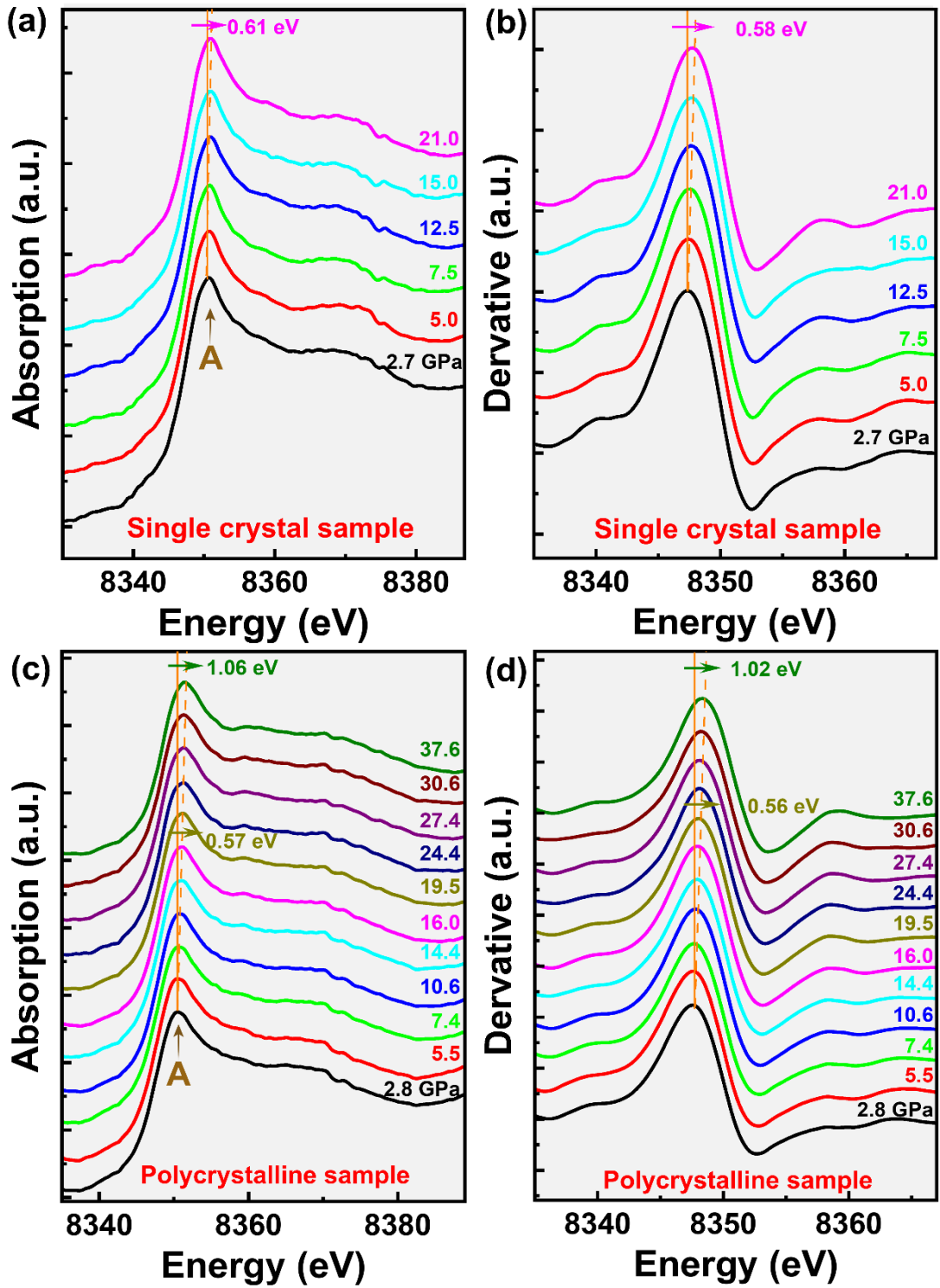

**Figure 2. Ni *K*-edge X-ray absorption spectra under different pressures.** (a) and (c) XAS data collected from the single crystal and polycrystalline $La_3Ni_2O_7$ samples, respectively. Observations of pressure-induced shifts in the white line peak (labeled by A) to high energies have been detected in both samples. (b) and (d) The corresponding

first derivatives of the XAS data in (a) and (c). The shifts of the white line peak in (b) and (d) to higher energies illustrate a gradual increase of the valence state of Ni ions with increased pressures.

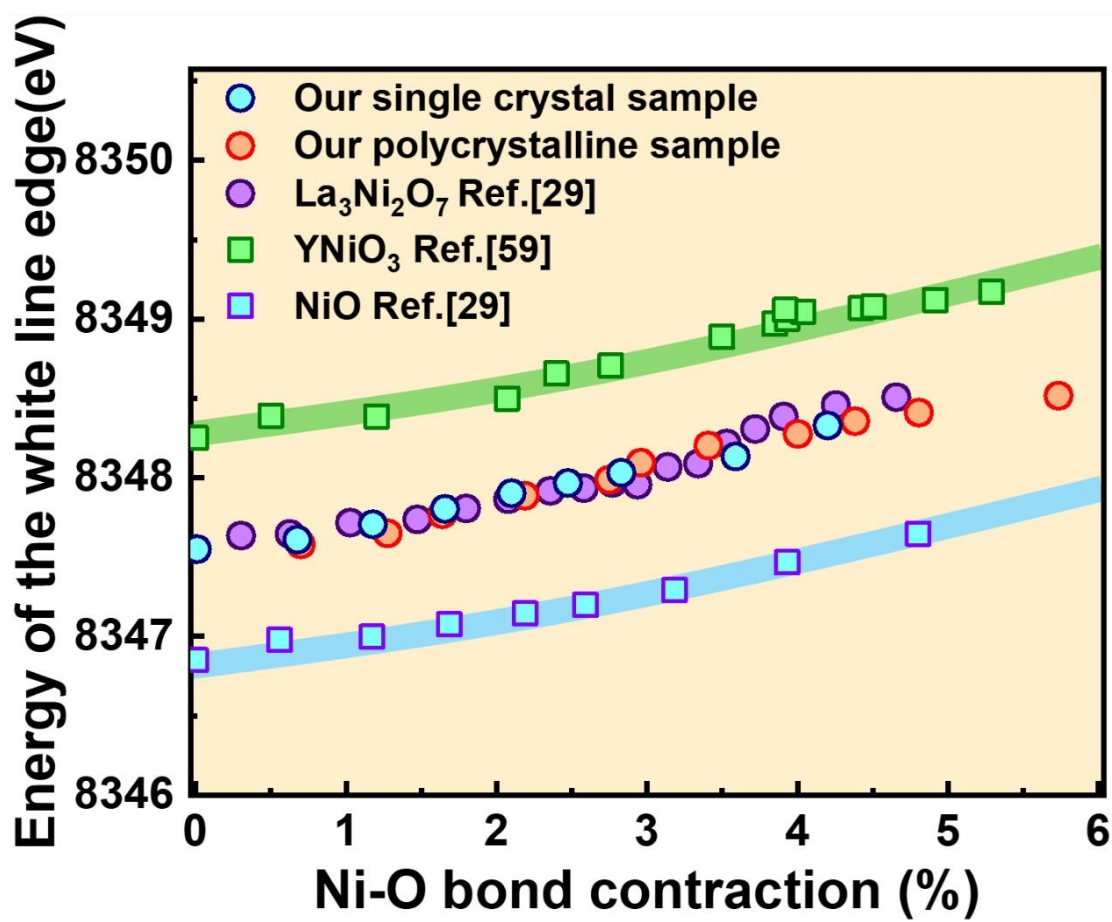

**Figure 3. Variation of energy edge of the white line peak with the contraction of the Ni-O bond measured from synchrotron X-ray absorption spectroscopy at different pressures.** The blue and orange circles are the data from our single crystal and polycrystalline $La_3Ni_2O_7$ ($Ni^{2.5+}$) samples. The purple circles are the $La_3Ni_2O_7$ data from Ref. [29]. The green and blue squares are the data from $YNiO_3$ ($Ni^{3+}$) and NiO ($Ni^{2+}$), which are taken from Ref. [59] and [29] respectively.

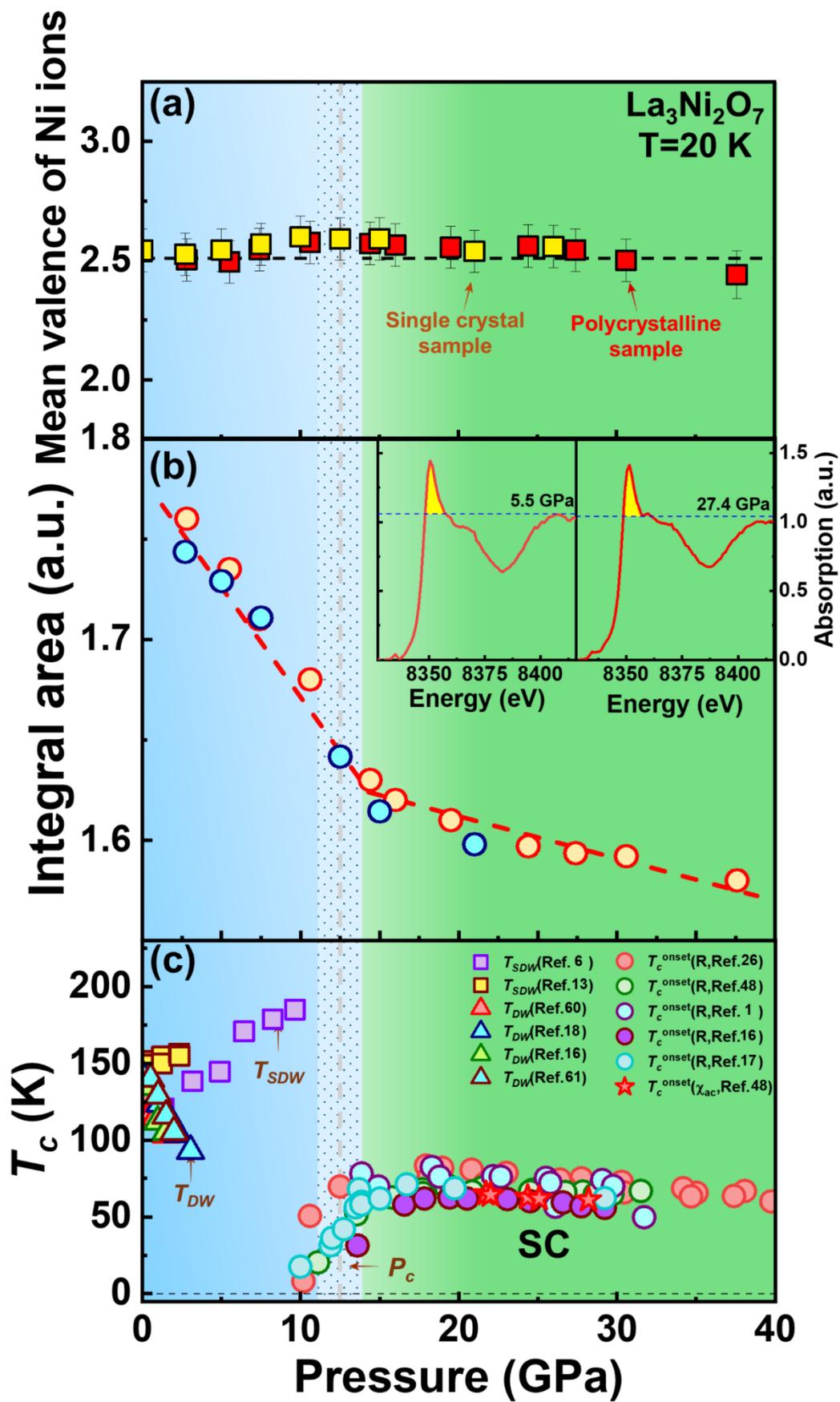

**Figure 4. Summary of pressure dependences of the low-temperature mean valence of Ni ions in La$_3$Ni$_2$O$_7$, white line peak area and transition temperatures of spin density wave (SDW), density wave (DW) and superconductivity (SC).** (a) The mean valence of Ni ions as a function of pressure. The red and yellow squares in the figure are derived from our measurements on the single crystal and polycrystalline samples, respectively. (b) The plot of the area of the white line peak versus pressure. The area is determined by integrating the highlighted region of the white line peak, as illustrated in the insets of the figure (b). (c) pressure-temperature phase diagram for La$_3$Ni$_2$O$_7$. $T_{\text{SDW}}$ and $T_{\text{DW}}$ in figure (c) stand for the formation temperature of the SDW and DW states. $T_c^{\text{onset}}$ ($R$) and $T_c^{\text{onset}}$ ($\chi_{ac}$) denote the $T_c$ value obtained from the resistance and modulated $ac$ susceptibility measurements.